\documentstyle[prl,aps,preprint,tighten]{revtex}

\newcommand{\beq}{\begin{equation}}
\newcommand{\eeq}{\end{equation}}
\def\bra#1{\langle #1 |} \def\ket#1{| #1 \rangle}
 
\def\expect#1{\langle #1 \rangle} \def\d{^\dagger} \def\L{{\cal L}}

\def\half{\frac{1}{2}}

\def\L{{\hat L}}

\begin{document}

\draft

\title{From quantum trajectories to classical orbits}
\author{T.A.~Brun,}
\address{Physics Department, Queen Mary and Westfield College,
University of London,\\
London  E1 4NS, England}

\author{N.~Gisin,}
\address{Group of Applied Physics,
University of Geneva, 1211 Geneva 4, Switzerland}

\author{P.F.~O'Mahony, M.~Rigo}
\address{Mathematics Department, Royal Holloway College,
University of London,\\
Egham, Surrey TW20 0EX, England}

\date{\today}

\maketitle

\begin{abstract}
Recently it has been shown that the evolution of open quantum systems
may be ``unraveled'' into individual ``trajectories,''  providing
powerful numerical and conceptual tools.  In this letter we use quantum
trajectories to study mesoscopic systems and their classical limit.
We show that in this limit,
Quantum Jump (QJ) trajectories approach a diffusive limit very
similar to the Quantum State Diffusion (QSD) unraveling.  The latter
follows classical trajectories in the classical limit.  Hence, both
unravelings show the rise of classical orbits.  This is true for both
regular and chaotic systems (which exhibit strange attractors).
\end{abstract}

\pacs{03.65.Bz, 03.65.Sq}

\section{Introduction}

Quantum mechanics is nonlocal. Classical mechanics is local.  Though it is
widely believed that quantum mechanics is the more fundamental theory,
attempts to describe classical phenomena by quantum equations are
fraught with difficulties.  Not only do the calculations become
extremely cumbersome, but they are conceptually more difficult.

There is no particular consensus on what it means to cross
the ``quantum $\rightarrow$ classical'' border.  Many criteria have been
suggested:  rapid decay of macroscopic superpositions, localization in
phase space, approach to coherent states, decoherence,
near-determinism, positivity of the Wigner
distribution, and non-violation of the Bell inequalities
\cite{JoosZeh85,Zurek91,DH,Spiller95,RigoGisin96}.  In fact,
it is rarely necessary to choose; as a rule, all of these are
satisfied for macroscopic systems.  Both modern experiments and the
emerging field of nanotechnology, however, are increasingly
challenging this divide.  As we probe the mesoscopic region,
where both quantum and classical effects are important, it behooves
us to have a better idea of what ``classical'' means.  This is not
only of theoretical interest; a better understanding should make it
possible to produce more efficient numerical models of mesoscopic systems.

In this letter, we show how the use of quantum trajectories, both
continuous and discontinuous, illuminates the intermediate
scales where neither a purely quantum nor a purely classical description is
practical.  These models not only simplify
computation (one of their major original motivations), but
describe when and how quantum nonlocality
disappears in the classical limit.
We illustrate our results by plotting regular and chaotic
quantum trajectories, and show the limit where they recover
classical regular and chaotic orbits.

The Schr\"odinger equation is the basic dynamical law of nonrelativistic
physics. However, strictly speaking, it applies only to the entire Universe
as the only truly closed system.
All other systems are open.  It is
well known that the environment is crucial for the emergence of classical
features in a quantum system. The best-known example is the case
of a measurement, leading to sharp values
of some physical quantity.  More generically, environments
induce decoherence, which is closely related to the rise of classical
properties \cite{JoosZeh85,Zurek91,DH}.
Consequently, we shall concentrate on open quantum
systems in the Markovian (time local) limit.

Markovian open quantum systems are usually
described by a master equation:
\beq
{\dot \rho} = - i [H,\rho] + \sum_m \biggl( L_m \rho L_m\d
  - {1\over2} \{ L_m\d L_m, \rho \} \biggr)
\label{master_eqn}
\eeq
where $\rho$ is the density matrix for the system, $H$ its Hamiltonian,
and the linear operators  $L_m$ describe the effects of the
environment. However, equation (\ref{master_eqn})
is not entirely satisfactory for our
purpose.  It describes only mean values,
computed over both quantum and classical probabilities. Hence,
density matrices by themselves do not tell us which features
can be described classically and which require a quantum description
\cite{Rigo96}.

Let us illustrates this distinction for the mean values over a density matrix
$\rho$ of the position operator $q$ and its square $q^2$.
These values give little idea
of the actual degree of localization of the particle.
A large value of the spread
$\left(Tr(q^2\rho)-Tr(q\rho)^2\right)^{\half}$ could correspond
either to delocalized particles (resulting from, e.g., the spreading
of a single wavepacket) or to localized particles whose
position is classically uncertain (resulting from, e.g., Brownian motion).
The evolution of localized particles can be efficiently
computed with classical or semi-classical models, while
delocalized particles necessarily require a less efficient but more
complete quantum description.

In the latter case, generally, there is nothing which corresponds
to classical phase space trajectories.
This is one of the chief difficulties in characterizing quantum chaos in a
way similar to that used in classical dynamics.

By {\it unraveling} the evolution of the density operator,
one obtains, as we shall see,
classical mixtures (i.e., classical probabilities) of quantum pure states
(i.e. quantum probabilities). This allows one to
distinguish quantum from classical and, at the same time, provides
a powerful tool for practical computations.

\section{Quantum State Diffusion and Quantum Jumps}

In such an unraveling, one describes the system in terms of a normalized
pure state $\ket{\psi(t)}$ which follows a stochastic ``trajectory'' in
Hilbert space.  By averaging the pure state projector $\ket\psi\bra\psi$
over all possible trajectories with appropriate weights, one reproduces
the density operator $\rho = M(\ket\psi \bra\psi)$.  This is analogous
classically to replacing the Fokker-Planck equation for probability densities
with a stochastic Langevin equation for single trajectories.

Unfortunately, unlike the case of classical Brownian motion, the unraveling
of the master equation (\ref{master_eqn}) is not unique.  Thus, there is
some ambiguity in how one separates classical and quantum uncertainties,
related to the ambiguity in identifying density matrices with quantum
ensembles.  In this section we consider two well known unravelings of
(\ref{master_eqn}).

In quantum state diffusion (QSD),
the (It\^o) stochastic evolution equation for the normalized state
vector $\ket{\psi(t)}$ reads:
\begin{eqnarray}
\ket{d\psi(t)} &=&  -i H \ket{\psi(t)} dt
  - \half\sum_j (L_j^\dagger L_j
  - 2\expect{L_j^\dagger}_\psi L_j
  + |\expect{L_j}_\psi|^2) \ket{\psi(t)} dt  \nonumber\\
&&   + \sum_j (L_j - \expect{L_j}_\psi) \ket{\psi(t)} d\xi_j
\label{QSDeq}
\end{eqnarray}
where the ``noises'' $d\xi_j$ are complex-valued Wiener processes of zero mean
$M(d\xi_j)=0$ and correlations
$M(d\xi_j d\xi_k)=0$, $M(d\xi_j^* d\xi_k)=\delta_{jk}dt$.
This equation describes a continuous non-differentiable
evolution similar to the familiar
diffusive paths of a classical Brownian particle, but in Hilbert space
instead of real space.  QSD is the only continuous unraveling which
satisfies the same symmetry properties as the master equation itself
\cite{GisinPercival}.

Our second example is the quantum jumps (QJ) unraveling, which
is closely related to photon counting. However, it can be
defined for any Lindblad master equation 
\cite{Dalibard92,Carmichael93}.
The stochastic increment for the wave-function is
\begin{eqnarray}
\ket{d\psi(t)} &=&  -i H \ket{\psi(t)} dt
  - \half\sum_j (L_j\d L_j
  - \expect{L_j\d L_j}_\psi) \ket{\psi(t)} dt  \nonumber\\
&& + \sum_j \biggl(
  \frac{L_j \ket{\psi(t)}}{\sqrt{\expect{L_j\d L_j}_\psi}}
  - \ket{\psi(t)} \biggr) dN_j
\label{QJeq}
\end{eqnarray}
The discrete Poissonian noises $dN_j$ assume the values 0 or 1.
Most of the time $dN_j=0$ and the evolution is continuous
and differentiable. However, whenever
$dN_j=1$ there is a ``jump'' to the state
$L_j \ket{\psi(t)}/\sqrt{\expect{L_j\d L_j}_t}$.
The $dN_j$ processes have mean values
$M_{\ket\psi}(dN_j)=\expect{L_j\d L_j}_\psi dt$ and correlations
$dN_j dt=0$ and $dNj dN_k=\delta_{jk}dN_j$.  This means, essentially,
that jumps occur randomly with an average rate $\expect{L_j\d L_j}$.

Let us illustrate these two unravelings for a simple example: the
damped harmonic oscillator at finite temperature. $H=\omega a^\dagger a$,
$L_1=\sqrt{\bar n\gamma}a^{\dagger}$ and
$L_2=\sqrt{(\bar n+1)\gamma}a$ where $\bar n$ is the thermal equilibrium
mean photon  number, $\bar n=\expect{a^{\dagger} a}_\rho$,
and $\gamma$ is the inverse relaxation time.
For QSD one can show that any initial states tends to a
coherent state: $\ket{\psi(t)} \rightarrow
\ket{\alpha_t}$, where $a\ket{\alpha_t} = \alpha_t \ket{\alpha_t}$
and $\alpha_t = (\expect{q}_\psi + i\expect{p}_\psi)/\sqrt(2)$.
Furthermore, the evolution
of $\alpha_t$ is governed by a classical equation,
\beq
d\alpha_t = -i\omega\alpha_t dt - {\gamma \over 2} \alpha_t dt
+ \sqrt{\bar n\gamma} d \xi_t
\label{dalpha}
\eeq
Hence, for this example at least, the QSD equation fully describes how the
environment localizes the quantum state down to a minimum Gaussian wavepacket,
and how this wavepacket follows a classical trajectory.

In \cite{GisinPercival,Percival94,Halliwell95} it was argued
that this is quite general, and not peculiar to the
harmonic oscillator. In \cite{SpillerRalph94} the QSD equation 
was applied to the Kaotic Anharmonic OScillator (KAOS)
system; in this case localization takes place despite
competition with the delocalizing nonlinear Hamiltonian.
In general, when the system remains small (in the sense that it does
not explore much of the phase space in units of $\hbar$),
the QSD trajectory presents no
definite structure.  However, when the parameters are such that the
system explores a larger portion of phase space, the quantum trajectories
exhibit a clear structure which approaches
the classical strange attractor.

Similar behavior has been demonstrated for other
chaotic systems, including the weak link capacitor 
and quantum kicked rotor \cite{Spiller95}, and
the forced damped Duffing oscillator \cite{Brun95}.  Once
again, as we approach the classical limit, the structure of the strange
attractor begins to appear.

The case of QJ is quite different. No result analogous to that for QSD holds
in the harmonic oscillator case. However,
in \cite{RigoGisin96} the same KAOS system was studied using the
QJ equation, and again the classical strange attractor was observed whenever
the system explored enough of phase space.

This result was very puzzling
at the time.  We can now show that this
behavior is also generic for QJ. In particular, we shall see that
whenever the system is far from the origin of phase space (i.e., the
harmonic oscillator ground state) its dynamics become very similar to QSD.

\section{Localization of QJ}

Consider a system in a state $\ket{\psi(t)}$
with a Hamiltonian $H$ and environment operators
$L_1 = \sqrt{\gamma_1} a$ and $L_2 = \sqrt{\gamma_2} a\d$.  Define
\begin{equation}
\ket{\phi(t)} = D(-\alpha_t)\ket{\psi(t)},
\end{equation}
where $\alpha_t=\expect{a}_{\psi(t)}$
and $D(\alpha)=\exp\{\alpha a^\dagger - \alpha^* a\}$
is the displacement operator.  $\ket{\phi(t)}$ is the state
$\ket{\psi(t)}$ displaced such that the mean position and momentum vanish:
$\bra{\phi(t)}a\ket{\phi(t)}=0$.
Note that $|\alpha_t|$ measures the phase space ``distance'' of the state 
$\ket{\psi(t)}$ from the origin $\ket0$. Let $\Delta\alpha_t$
measure the width of the state $\psi_t$:
\beq
\Delta\alpha_t^2 \equiv \expect{a^\dagger a}_\psi -
\expect{a^\dagger}_\psi \expect{a}_\psi =
\bra{\phi(t)} a^\dagger a \ket{\phi(t)}.
\eeq
Now consider the following limit.  When the oscillator is
well localized relative to its distance from the origin, so that
$|\alpha_t| >> \Delta\alpha_t$, then the rate of jumps $\expect{L^\dagger L}_t
\approx |\alpha_t|^2$ is large, while the size of the jumps are
small: $\bra{\psi(t)}(a\d - \alpha_t^*)(a - \alpha_t) \ket{\psi(t)} =
\Delta\alpha_t$, and similarly for $a\d$.
Hence, when the energy of the system is large relative to
$\hbar$, the ``jumpy'' evolution approaches closer and closer to
a diffusion process like QSD, describing localized wavepackets
following classical trajectories.

Let us develop the QJ equation (\ref{QJeq}) to first order in
$\Delta\alpha/|\alpha|$. To remove irrelevant phases, we use
the 1-dimensional projector $P_t \equiv \ket{\psi(t)}\bra{\psi(t)}$. 
From equation (\ref{QJeq}) one obtains:
\begin{eqnarray}
dP_t &=& - i [H,P_t] dt + \sum_j \left(-\half\{L_j\d L_j, P_t\}
  + \expect{L_j\d L_j}_\psi P_t \right) dt \nonumber\\
&& + \left(\frac{L_j P_t L_j\d}{\expect{L_j\d L_j}_\psi}
  - P_t \right) dN_j.
\label{Projector}
\end{eqnarray}

Using $a\ket\psi = \alpha\ket\psi + D(\alpha)a\ket\phi$
and $a P_t a^\dagger=|\alpha_t|^2 P_t + \alpha_t^*(a-\alpha_t)P_t
+ \alpha_t P_t (a^\dagger-\alpha_t^*)+ O^2(\frac{\Delta\alpha}{|\alpha|})$,
one deduces:
\beq
\frac{aP_t a^\dagger}{\expect{a^\dagger a}_t}
  = P_t + \frac{a-\alpha_t}{\alpha_t} P_t +
  P_t \frac{a^\dagger-\alpha^*_t}{\alpha_t^*}
  + O\left(\frac{\Delta\alpha^2}{|\alpha|^2}\right),
 \label{FirstOrdeq}
\eeq
and similarly for $a\d P_t a$.  (These are our only environment operators.)
Inserting this into (\ref{Projector}) yields
\begin{eqnarray}
dP_t &\approx& - i [H,P_t] dt + \sum_j \left( L_j P_t L_j\d
  - \half\{L_j\d L_j, P_t\} \right) dt \nonumber\\
&& + (L_j - \expect{L_j}_\psi) P_t {\expect{L_j\d}\over|\expect{L_j}|} dW_j
  + P_t (L_j\d - \expect{L_j\d}_\psi) {\expect{L_j}\over|\expect{L_j}|} dW_j,
\label{dP}
\end{eqnarray}
where we have made the approximation
\begin{equation}
{\expect{L_j\d}\over\sqrt{\expect{L_j\d L_j}}}
  \left( {{dN_j}\over{\sqrt{\expect{L_j\d L_j}}}}
  - \sqrt{\expect{\L_j\d L_j}} dt \right)
  \approx \frac{\expect{L_j\d}}{|\expect{L_j}|} dW_j,
\end{equation}
with the $dW_j$ standard real Wiener
processes ($dW_i dW_j = \delta_{ij}dt$).  Hence,
\begin{eqnarray}
\ket{d\psi(t)} &\approx&  -i H \ket{\psi(t)} dt
  - \half\sum_j (L_j^\dagger L_j
  - 2\expect{L_j^\dagger}_\psi L_j
  + |\expect{L_j}_\psi|^2) \ket{\psi(t)} dt  \nonumber\\
&& + \sum_j (L_j - \expect{L_j}_\psi) \ket{\psi(t)}
  {\expect{L_j\d}\over|\expect{L_j}|} dW_j.
\label{QJlimiteq}
\end{eqnarray}

Note that this equation (\ref{QJlimiteq})
preserves the norm of $\ket{\psi(t)}$ and
recovers (\ref{master_eqn}) in the mean:  $\rho(t) =
M(\ket{\psi(t)}\bra{\psi(t)})$ at all times.
It is almost identical to the QSD equation
(\ref{QSDeq}), except that the noise is a real Wiener process multiplied
by a phase which depends on the phase space position of the
oscillator.  This is the only choice of phase such that the state vector
remains normalized in the diffusive limit.  The QSD equation was chosen
to be independent of the choice of phase.
The solutions of equations (\ref{QSDeq})
and (\ref{QJlimiteq}) are very similar, as we shall see.

We can see this by going to the classical limit
and plotting the trajectories for the QJ equation.  We have
done this for an extension of the thermal model presented in the previous
section.  Note how in this case the jumps occur very frequently as we move
away from the origin of phase space.  (See figure 1.)  This also works for
the chaotic cases already studied with QSD; in the QJ
case as well, we can see the emergence of classical structure.
(See figure 2.)

A similar derivation can be done for any
choice of environment operators $L$ which are linear in $a$ and $a\d$
(e.g., $x$ and $p$).

\section{Conclusion}

We have seen how the use of stochastic evolution equations
for pure states provides the ability to draw useful pictures of
elementary quantum phenomena.  These same equations provide powerful
tools for practical numerical computations; in particular, one can
exploit the existence of localized solutions to reduce the numerical
difficulty of solving the equations, using the technique of {\it moving
bases} \cite{Schack95,Steimle95,Holland96,Schack96} to produce more
efficient computer simulations.  Quantum unravelings
have long suggested the possibility of separating
quantum and classical uncertainties for open systems; however, the
ambiguity in the choice of unraveling has prevented any conclusions
from being drawn as to the exact meaning of this separation.

We now see that, as one approaches the classical limit, it should be
possible to make this separation in a similar way for different
unravelings.  This suggests that, in a sense, the details don't
matter:  there is a single classical limit towards which a broad
class of different ``quantum trajectory'' techniques all tend.  In treating
physical systems in the mesoscopic regime, this should make it possible
to determine unambiguously which characteristics may be given
classical and which quantum descriptions.  This, in turn, may contribute
greatly to the numerical solution of practical problems.

We should also stress the limitations of this approach.  In the first
place, it depends on the ``system/environment split,'' which will always
include a certain ambiguity.  While the exact boundary between system and
environment is not important for macroscopic systems, for a microscopic
or mesoscopic system it can be crucial.  Second, we have limited ourselves
to the Markovian approximation.  This is mainly because little is known
about the more general case.  We expect that similar considerations apply
to non-Markovian systems.

We would like to thank Lajos Di\'osi, Francesca Mota-Furtado,
Ian Percival and R\"udiger Schack for useful conversations.  This
research was funded in part by the UK EPSRC, the EU Human Capital and
Mobility Programme, and the Swiss National Science Foundation.

\vfil

Figure 1.  The harmonic oscillator with $\omega=1$ at finite temperature,
  the equilibrium displaced by a constant
  force $\beta$; curves are plotted for the values
  $\beta=1,4,10$ and average thermal excitation ${\bar n} = 0.2$.
  Each jump is marked; as the oscillator is displaced further from the
  origin, the jumps become frequent and small in effect, illustrating
  the transition from a ``jumpy'' trajectory to a diffusion process.

\vfil

Figure 2.  The Poincar\'e section of the forced damped Duffing oscillator
  in the chaotic regime.  The value $\beta$ gives the scale of the system,
  with $\beta \rightarrow \infty$ the classical limit; this system classically
  exhibits dissipative chaos, and has already been investigated in this limit
  using QSD (see \cite{Brun95}).  We keep $\hbar = 1$ constant.
  As the scale increases for $\beta=1,4,10$,
  the structure of the strange attractor clearly emerges, showing the
  diffusive limit of quantum jumps.  This is similar to the behavior
  observed for QSD.  The classical result is included for comparison.

\vfil

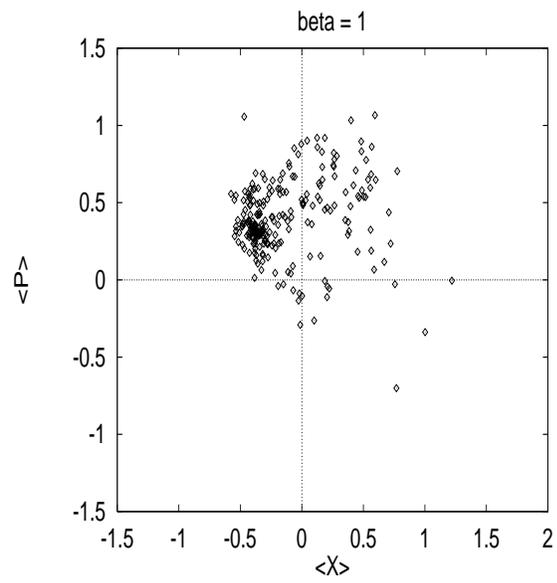
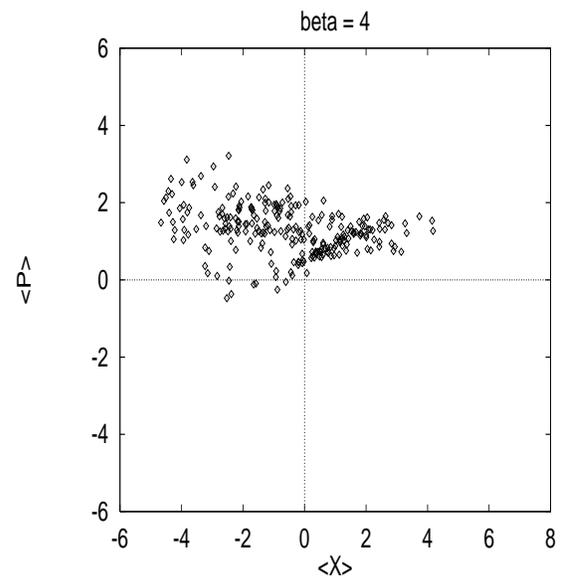
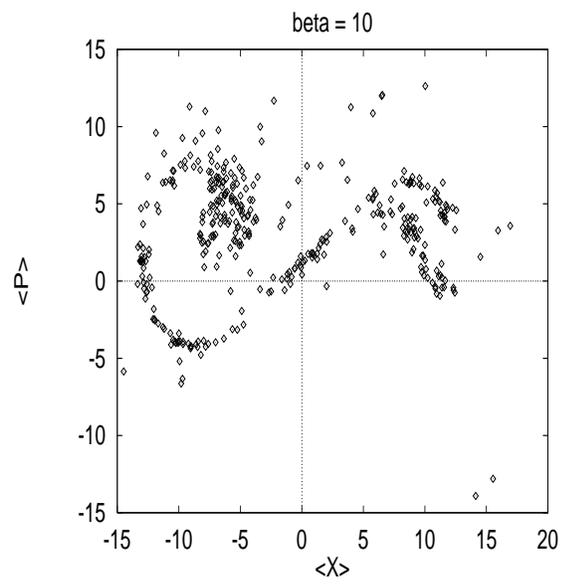
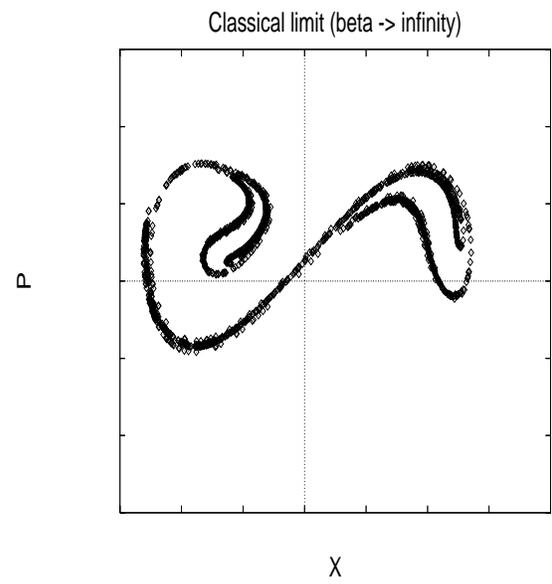
\end{document}